\documentclass[final,5p,times,twocolumn]{elsarticle}

\usepackage{amssymb}
\usepackage{gensymb}

\usepackage[dvipsnames]{xcolor}

\journal{Applied Radiation and Isotopes}

\begin{document}

\begin{frontmatter}



\title{The i-TED Compton Camera Array for real-time boron imaging and determination during treatments in Boron Neutron Capture Therapy}


\author[inst1]{Pablo Torres-Sánchez}
\author[inst1]{Jorge Lerendegui-Marco}
\author[inst1]{Javier Balibrea-Correa}
\author[inst1]{Víctor Babiano-Suárez}
\author[inst1]{Bernardo Gameiro}
\author[inst1]{Ion Ladarescu}

\author[inst2,inst3]{Patricia Álvarez-Rodríguez}
\author[inst3]{Jean-Michel Daugas}
\author[inst3]{Ulli Koester}
\author[inst3]{Caterina Michelagnoli}
\author[inst4]{María Pedrosa-Rivera}
\author[inst4]{Ignacio Porras}
\author[inst5]{Mª José Ruiz-Magaña}
\author[inst6]{Carmen Ruiz-Ruiz}

\author[inst1]{César Domingo-Pardo}

\affiliation[inst1]{organization={Instituto de Física Corpuscular (CSIC-Universidad de Valencia)},
            city={Valencia},
            country={Spain}}

\affiliation[inst2]{organization={Instituto de Biopatología y Medicina Regenerativa, Centro de Investigación Biomédica. Universidad de Granada},country={Spain}}

\affiliation[inst3]{organization={Institut Laue-Langevin},
            city={Grenoble},
            country={France}}

\affiliation[inst4]{organization={Departamento de Física Atómica, Molecular y Nuclear, Universidad de Granada},country={Spain}}

\affiliation[inst5]{organization={Departamento de Biología Celular, Universidad de Granada},country={Spain}}

\affiliation[inst6]{organization={Departamento de Bioquímica y Biología Molecular III e Inmunología, Universidad de Granada},country={Spain}}

\begin{abstract}

This paper explores the adaptation and application of i-TED Compton imagers for real-time dosimetry in Boron Neutron Capture Therapy (BNCT). The i-TED array, previously utilized in nuclear astrophysics experiments at CERN, is being optimized for detecting and imaging 478 keV gamma-rays, critical for accurate BNCT dosimetry. Detailed Monte Carlo simulations were used to optimize the i-TED detector configuration and enhance its performance in the challenging radiation environment typical of BNCT. Additionally, advanced 3D image reconstruction algorithms, including a combination of back-projection and List-Mode Maximum Likelihood Expectation Maximization (LM-MLEM), are implemented and validated through simulations. Preliminary experimental tests at the Institut Laue-Langevin (ILL) demonstrate the potential of i-TED in a clinical setting, with ongoing experiments focusing on improving imaging capabilities in realistic BNCT conditions.

\end{abstract}

\end{frontmatter}


\section{Introduction}

In BNCT real time dosimetry of a patient is key for planning and monitoring a treatment. By now, the standard method uses the boron concentration measured in blood, using  inductively coupled plasma (ICP) techniques, coupled with a 18F-FBPA PET scan performed on the patient prior to the treatment, in order to estimate the boron distribution. However, this poses severe limitations to the accuracy of the dosimetry, as it includes extrapolations from the labelled compound on a different date, and assumes the same pharmacokinetic profile for the blood and the tumor and organs of interest. As such, the measurement of boron concentration during treatment still represents a major challenge for BNCT \cite{IAEA_TECDOC}. 

In fact, boron concentration determination alone for BNCT dosimetry is not sufficient, as it needs to be multiplied by the neutron fluence at each point, thus increasing the overall uncertainties. However, neutron capture reactions in boron can be tagged by means of the emission of a 478 keV gamma-ray, which can be subsequently used to monitor the product of both the boron concentration and neutron fluence. This quantity would therefore enable an accurate assessment of the boron dose, which is the main component of the dose in a BNCT treatment. In addition, this is proportional to the thermal neutron dose, due to the similar behavior of the $^{14}$N($n,p$) reaction (main component of the thermal dose) and neutron capture in hydrogen, which produces gamma-rays with an energy of 2224 keV. Therefore, a better estimation of the thermal neutron dose could be also achieved if these 2224 keV gamma-rays were imaged. 

The use of the 478 keV radiation to track the total dose received during a BNCT treatment has been proposed in several previous studies \cite{Verbakel97}. The most basic approach is to just measure the 478 keV gamma-rays using a spectrometer located close enough to the patient, and with resolution high enough to separate it from 511 keV annihilation gamma-rays. 

However, these techniques do not allow to spatially discriminate if the location of high dose deposition is spread over a large region, or to properly separate the dose received by the tumor and that of surrounding tissues. The latter component is expected to be lower due to the reduced local boron concentration, but it may generate a similar, if not larger, 478~keV gamma-ray contribution due to the overall boron content in healthy tissues. Moreover, the surrounding walls of the treatment room are covered with borated shielding to absorb neutrons for radioprotection, which further increases a background that cannot be easily subtracted from the actual tumor-originated contribution. Some improvements of this strategy include opening a hole in the walls of the treatment room and use it as a collimator to improve the selectivity to the tumor region and reduce the ambient background \cite{Sakurai09}.

In order to capture the full boron dose distribution, including the spatial distribution in the tumor and surrounding organs at risk, and also measure variations of such dose during the extent of the treatment, a real-time imaging technique is needed. Two main strategies have been proposed, one based on single photon emission computed tomography (SPECT) \cite{Minsky11,Manabe16,Fatemi19,Silarski23,Caracciolo23} and the other on Compton Cameras (CC) \cite{Lee15, Gong18}. Both techniques are based on the same concept of gamma-ray imaging. In principle, SPECT is able to generate high resolution images, due to the large mechanical collimation system placed in between the imaging volume and the gamma-ray detectors. However, this intrinsically reduces the efficiency by a large factor, which may be a drawback for real-time monitoring. Additionally, the collimation is not complete for intermediate-energy (478 keV) gamma-rays, which in turn requires extra thickness for the collimators. Finally, since the imaging system is to be placed inside the treatment room near the patient, the large neutron fluence generates a high radiation background that is expected to lower the signal-to-background ratio for the pursued application. Also, given the large neutron fluence, a non-negligible activation builds up during the treatment. Both for lead and tungsten collimators neutron-activation may lead to radioprotection issues. In contrast, CCs offer more compact systems with high efficiency and seem to be more compatible with the clinical BNCT environment. Moreover, the modularity of these systems allows the use of several of them or to install them in moving frames to enable the possibility of 3D tomographic reconstruction, while keeping the overall equipment volume relatively low. 
Several CC designs and concepts have been proposed thus far for BNCT. The latter include the use of different radiation-sensitive materials, such as semiconductors HPGe \cite{Stockhausen12}, CdZnTe \cite{Kim23}, Si/CdTe \cite{Sakai23}, or scintillator crystals of LaBr$_3$ \cite{Nutter24}. Regarding Compton-image reconstruction, several options have been already studied, both for BNCT and other purposes \cite{Tashima22}. Specially relevant is the use of iterative procedures such as Maximum Likelihood Expectation Maximization (MLEM), Ordered Subset Expectation Maximization (OSEM) \cite{Ramos23}, and neural networks (e.g. GANs \cite{Hou22}).

So far, the main challenges in most previous studies are related to attaining the spatial resolution required, true online capabilities and dealing with the harsh radiation backgrounds induced by the neutron-beam during treatment.


In this work we report on the possibility to adapt and apply i-TED Compton imagers, which are readily available from nuclear-astrophysics experiments carried out at CERN n\_TOF~\cite{hymns,Domingo16,Babiano21, Lerendegui22_NIC, Lerendegui23}. Empowered by the specific optimization-design and the excellent performance of these imagers in the harsh neutron-induced background conditions~\cite{Babiano20,Balibrea21}, in this work we explore and quantify the prospects of these gamma-cameras for dosimetry in BNCT.

\begin{figure}[h!]
    \centering
    \includegraphics[width=0.45\textwidth]{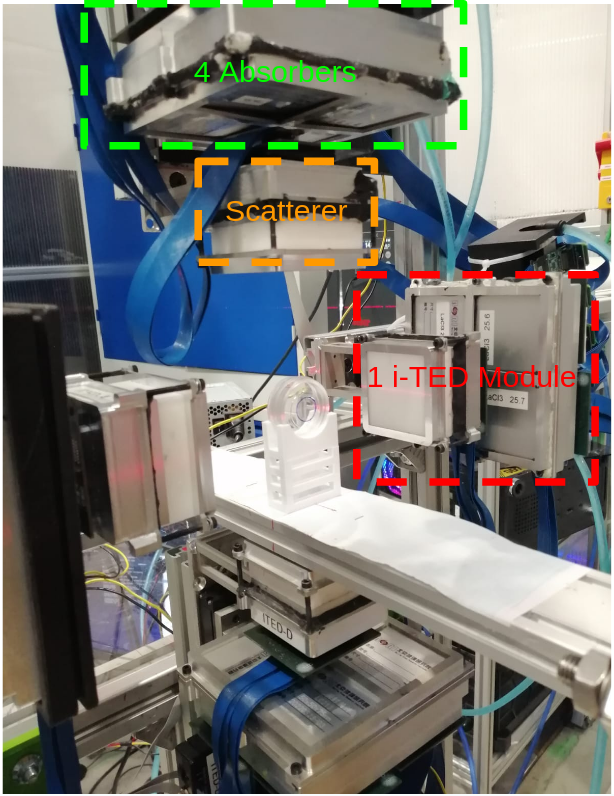}    

    \caption{Picture of the full i-TED array with four modules arranged in cross-shape. Each i-TED module consists of one scatterer and four absorber crystals, as indicated in the picture.}
    \label{fig:iTED_description}
\end{figure}

The i-TED array consists of four Compton modules, each of them including one scatterer crystal and four absorber crystals. This original design with a four-fold absorber size with respect to the scatterer detector becomes of particular interest for enhancing detection efficiency at relatively low (478 keV) gamma-ray energy owing to the energy-dependency of the Compton effect~\cite{Babiano20}.
All i-TED crystals are monolithic LaCl$_3$(Ce) scintillators of 5$\times$5 cm² cross-sectional area. The scatterers have a thickness of 1.5 cm, while the absorbers are 2.5 cm thick. The aforementioned 1:4 S:A aspect-ratio allows to achieve a high coincidence efficiency, which is specially important for low energy gamma-rays, as for instance, for 478 keV gamma-rays of initial energy, a 100 keV energy deposition in the scatterer, close to the detection limits, is followed by a gamma-ray emitted with an angle of around 45\degree. 
Within each detector, the scintillation photons are recorded by means of silicon photomulipliers or SiPMs (8x8 pixels, SensL ArrayJ-60035-64P-PCB), which are readout via PETSys-based TOF-PET ASIC electronics. A 10 Gigabit link is used to transfer data to the adquisition computers. In order to minimize thermal-gain fluctuations in the detectors a liquid-cooling system has been also implemented. Figure \ref{fig:iTED_description} shows the i-TED array in cross-shaped configuration.

The article is structured as follows. The materials and methods section (Sec. \ref{sec:MandM}) describes the Monte Carlo (MC) simulations used to characterize the expected capabilities of i-TED towards BNCT dosimetry and to optimize or adapt its features for 478 keV gamma rays. This includes also a description of recent developments in 3D image reconstruction algorithms carried out in our group. The results and discussion section Sec. \ref{sec:Results_DetUpdates} discusses the proposed detector upgrades and Sec. \ref{sec:Results_3Drec} summarizes first tests of the 3D image reconstruction algorithms based on MC-simulations. Finally, Sec. \ref{sec:prospects} describes the future prospects and introduces recent proof-of-concept experiments performed at Institut Laue-Langevin (ILL) where the full array of i-TED cameras in its present status have been tested. Finally, the main results and conclusions of this work are summarized in Sec. \ref{sec:Conclusions}.

\section{Materials and Methods}
\label{sec:MandM}
A series of Monte Carlo simulations have been carried out to explore the radiation conditions and imaging capabilities of the i-TED Compton Camera array for dosimetry in BNCT. The final step in the set-up optimization for BNCT will include the simulation of the detector response with anthropomorphic (ICRP-110) \cite{ICRP110} and simplified phantoms (Snyder model) \cite{medicalmodelsmcnp} in conjunction with the neutron beam coming out of the BSA developed within the NeMeSis project \cite{TorresSanchez21} to test the radiation conditions and background levels. These two scenarios have been already implemented in this work, thereby including typical boron concentrations of 20 ppm in healthy tissues and 65 ppm at the tumor regions. 
MCNP6.3 \cite{MCNP63} was used for simulating the neutron source, neutron transport and the production of secondary gamma-rays (specifically 478~keV and 2224 keV radiation) in the phantoms. \textsc{Geant4}\cite{Geant4} was used in a second simulation stage, which included neutron- and gamma-transport, as well as interaction with the detector and event identification. MCNP and \textsc{Geant4} simulations were streamlined through at pipeline based on the MCPLtools \cite{MCPL}, profiting from the SSW feature of MCNP to track and record neutrons and gamma-rays crossing a defined surface, and then the G4MCPLGenerator class in \textsc{Geant4} to use those same particles as primary generator action. 

However, in a first stage of detection-setup optimization and development of suitable image-reconstruction algorithms a simplified MC simulation was carried out. The latter essentially consisted of point-like gamma-ray sources at different places in the detectors field of view, as discussed later in Sec.~\ref{sec:Results_3Drec}.
Indeed, in order to obtain dosimetric data from the measurements a suitable image reconstruction algorithm needs to be implemented. In BNCT, given the sizes of both the tumor and surrounding tissues of interest, which can have relevant boron uptakes, 3D reconstruction is highly convenient \cite{IAEA_TECDOC}. A single Compton Camera may provide 3D image reconstruction although with rather limited performance in terms of resolution in the normal direction \cite{Ramos23}. Therefore, a possible solution consists of 3D tomography by means of a detectors array located around the imaging volume with different orientations. Alternatively, one could also utilize a single camera mounted on a moving structure and taking a series of measurements from different orientations. In the present study, a four-orientation in a cross-shaped configuration was used for the 3D image reconstruction, as shown in Fig.~\ref{fig:iTED_description}.

Given the various techniques for image reconstruction, one can think of using well-developed 2D image reconstruction algorithms, and stack them in 3D by taking the shots at several planes. This, however, introduces the limitation of the possible orientations, since all of them need to be mutually perpendicular, or they may need posterior interpolation which reduces resolution. Thus, a pure 3D algorithm is needed, which allows the reconstruction irrespective of the initial location of the camera (i.e. positions of the coincidence detections in a pair of crystals).

The proposed algorithm for this task is 1) the use of a 3D back-projection algorithm, that includes the position- and energy-resolution into the cone reconstruction \cite{Tashima20}, and 2) a List-Mode Maximum Likelihood Expectation Maximization (LM-MLEM) algorithm that uses the 3D back-projection output as the initial guess for the iterative procedure and allows the unfolding of the gamma-ray source distribution \cite{VojtechPoritz23}. These algorithms include a model function that analytically computes the probability of any event in the list being originated in each voxel. In addition, the 3D sensitivity matrix that is needed for LM-MLEM can be precomputed by means of Monte-Carlo simulations. This is especially interesting as this sensitivity matrix can be tuned to incorporate attenuation corrections that relieve the effect of gamma-ray absorption and scattering inside the human tissues (or any other material present in the imaging region). This can be straightforwardly achieved by extracting the attenuation data from previous CT-scans performed to the patient in preparation for the treatment planning. 

In general, these algorithms are slow and have a strong demand of resources in terms of RAM. In addition, a large amount of events is needed in order to reconstruct an image (e.g. 10k coincident Compton events are needed to properly reconstruct a point source). Furthermore, the intense radiation field and gamma-ray production inside the patient can produce count rates of the order of 50k coincidence events per second in each module, which can easily overwhelm the processing procedure. For these reasons, intensive computing strategies are needed. One such technique is the use of parallel, multithreading and GPU-based computations. One fitting tool for this is the SYCL programming model, whose implementations allow the same codes to be used in several different hardware accelerators, including CPU, GPU and FPGAs, from various vendors in the market. To such goal, our team has already developed a parallel algorithm to compute the back-projection (i.e. the first step of the 3D reconstruction algorithm)  \cite{Gameiro24}. This implementation has been tested in several computers and architectures, including CPU (Intel Core i7-1185G7, Intel Core  i9-14900HX) and GPU (NVIDIA GeForce RTX 4080 Laptop, NVIDIA GeForce GTX 1650Ti, Intel Iris Xe Graphics Tigerlake-LP GT2, Intel Data Center GPU Max 1100). Currently, this approach allows to process up to 70k Compton events in a single processing batch, with computing times below 8.25 s for 50k Compton events using a laptop with a GPU NVIDIA GeForce RTX 4080.

The second step, LM-MLEM, critically depends on the system model used to compute probabilities, and the way it is implemented into the iterative algorithm. Precomputation of this matrix allows a much faster iteration, and thus reaches 200 iterations in few minutes instead of few hours that takes when the matrix is recomputed after each iteration, as that computation takes 97\% of the total computing time of a given iteration. This advantage in time comes at the expense of the number of events that can be included in the iterative process (or alternatively the number of voxels in the imaging volume, i.e. the spatial resolution). For a 32 GB RAM and using float variables of single precision, this limit appears around 30k events with a 20$\times$20$\times$20 cm³ voxelized volume and 0.5~cm resolution. This number of events might be enough for spotting one or few point sources within the imaging volume, but ideally should be improved in order to uncover a spread distribution as expected in a BNCT treatment. Current efforts in reducing the computing time lie on parallelizing the system-model matrix with the aim of mitigating the limits imposed by precomputation in terms of resources.

\section{Results and discussion}

\subsection{Detector upgrades}
\label{sec:Results_DetUpdates}
One of the key elements for adapting i-TED to BNCT is the $\gamma$-ray detection efficiency. In particular, i-TED was optimized for detecting Compton events from gamma radiation of intermediate energy around 1-2~MeV and, therefore, the geometry could be further optimized for detecting 478~keV gamma-rays.

Neutron fluxes and subsequent gamma-ray production rates in BNCT are high enough so that the detection efficiency is not the limiting factor in terms of statistics. Quite on the contrary, the very high counting rates at the detectors pose strong constraints to the use of online dose monitors close to the patient.

Under these circumstances, it becomes convenient to reduce the overall single event detection efficiency (i.e. per scintillator) by reducing the thicknesses of the crystals in the current i-TED version. Moreover, one can optimize the thicknesses of both scatterer and absorber crystals so that the total coincidence efficiency for Compton imaging at the relevant gamma-ray energy does not decrease significantly. This is of great advantage to the monitoring, as the Compton event statistics can be preserved even when the incoming radiation flux is higher, thus allowing to bring the detector closer to the patient, and therefore improving significantly the spatial resolution (given a fixed angular resolution).

Incidentally, by reducing the crystal thicknesses, the event reconstruction of each gamma-ray absorption or scattering inside any crystal results in a better position resolution and hence improves the final resolution of the reconstructed images \cite{Balibrea21}.

\begin{figure}[h!]
    \centering
    \includegraphics[width=0.50\textwidth]{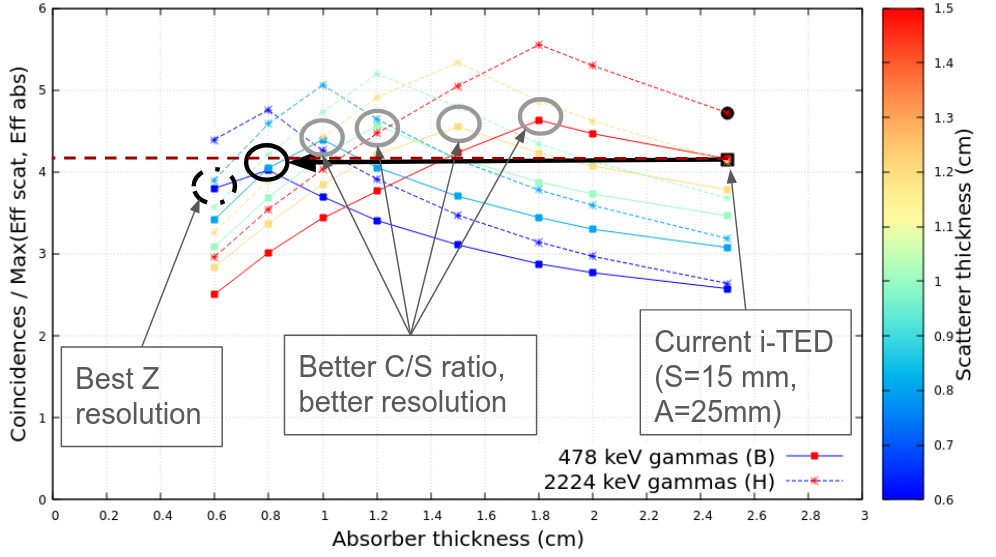}
    \caption{Optimization of the scatterer and absorber thicknesses. Solid lines are used for 478 keV gamma-rays (used for the optimization), and the values for 2224 keV gamma-rays are also given in dashed lines (for comparison). The absorber thickness is indicated in the X axis, and the scatterer thickness is marked with the color code. Coincidences to total efficiency ratio is used for the optimization.}
    \label{fig:ScintillatorThicknessOptimization}
\end{figure}

Figure \ref{fig:ScintillatorThicknessOptimization} shows the ratio of the Compton coincidence events to the maximum of single-detection events either in scatterer or absorber (i.e. the one with the maximum and thus limiting count rate), for different thicknesses of both the scatterer (color coded, from 6 mm in blue to 15 mm in red) and absorber (X axis, from 6 mm to 25 mm). The original i-TED configuration (15 mm scatterer, 25 mm absorber) is indicated for reference with the large black marker. Solid lines correspond to the Compton coincidences from 478 keV gamma-rays, and dashed lines are referring to 2224 keV gamma-rays from capture in hydrogen. It can be clearly noticed that there is a series of combinations of scatterer-absorber thicknesses that maintain a similar Compton coincidence efficiency ratio, with no room to a much larger improvement of this figure. However, the fact that reducing the crystal thickness improves both the position-reconstruction and allows higher incident fluxes without raising the count rates tilts the choice towards lower thicknesses for both scatterer and absorbers. Values of 6 mm for the scatterer and 8 mm for the absorbers are found as a good compromise for all the parameters discussed in this optimization procedure.


\subsection{3D image reconstruction}
\label{sec:Results_3Drec}
\begin{figure*}[h!]
    \centering
    a)
    \includegraphics[width=0.45\textwidth]{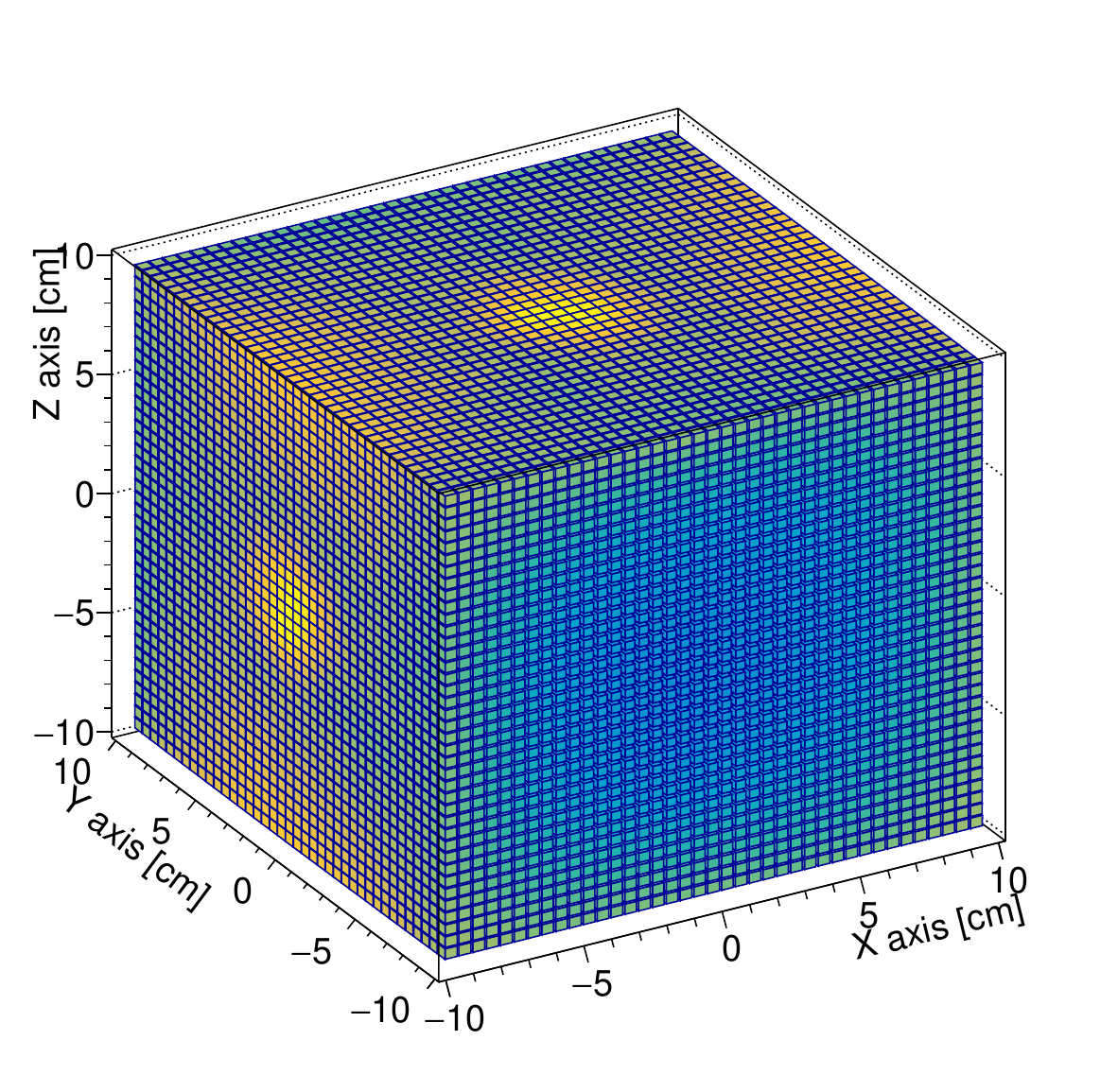}
    b)
    \includegraphics[width=0.45\textwidth]{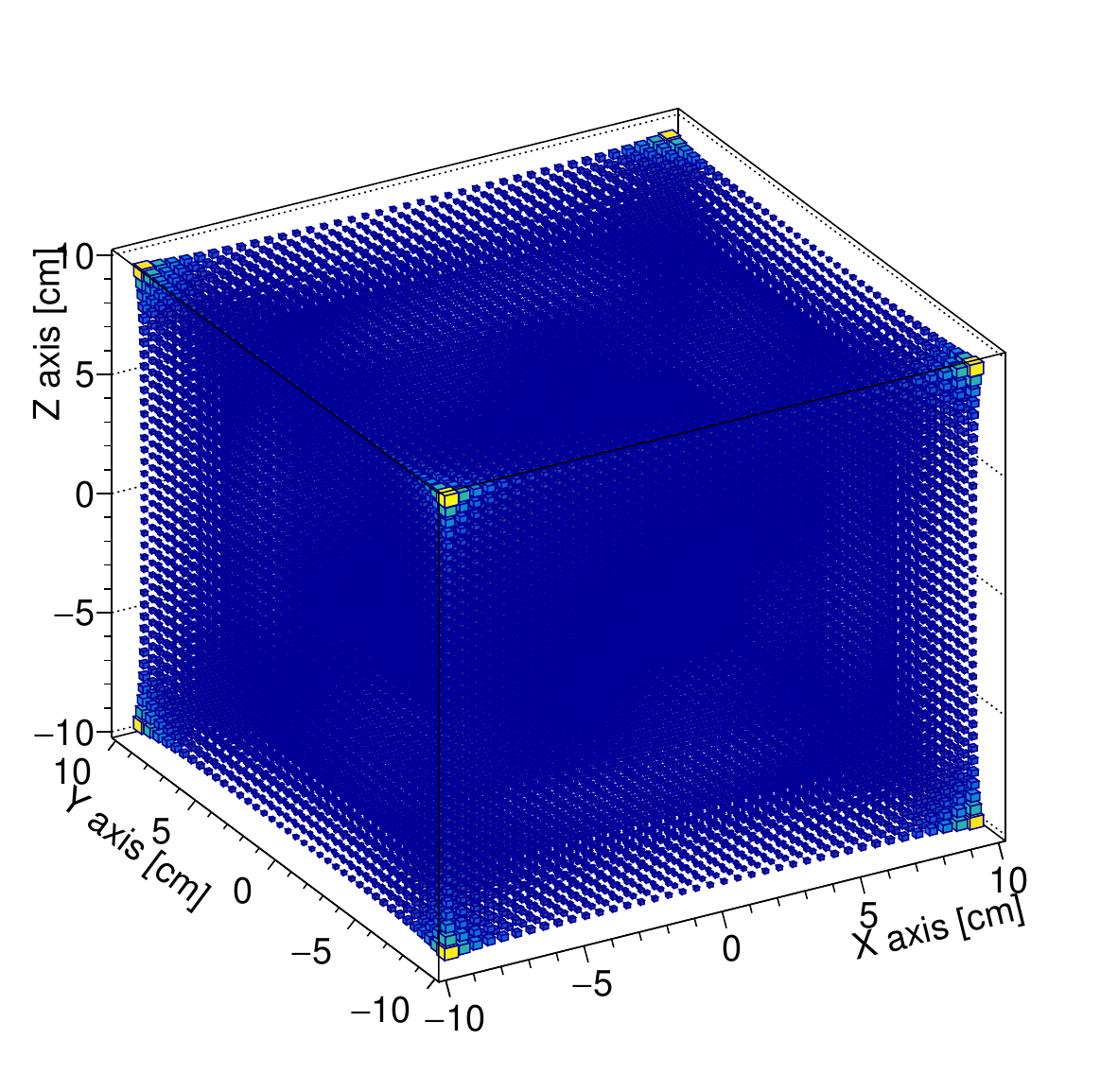}
   \\ c)
    \includegraphics[width=0.45\textwidth]{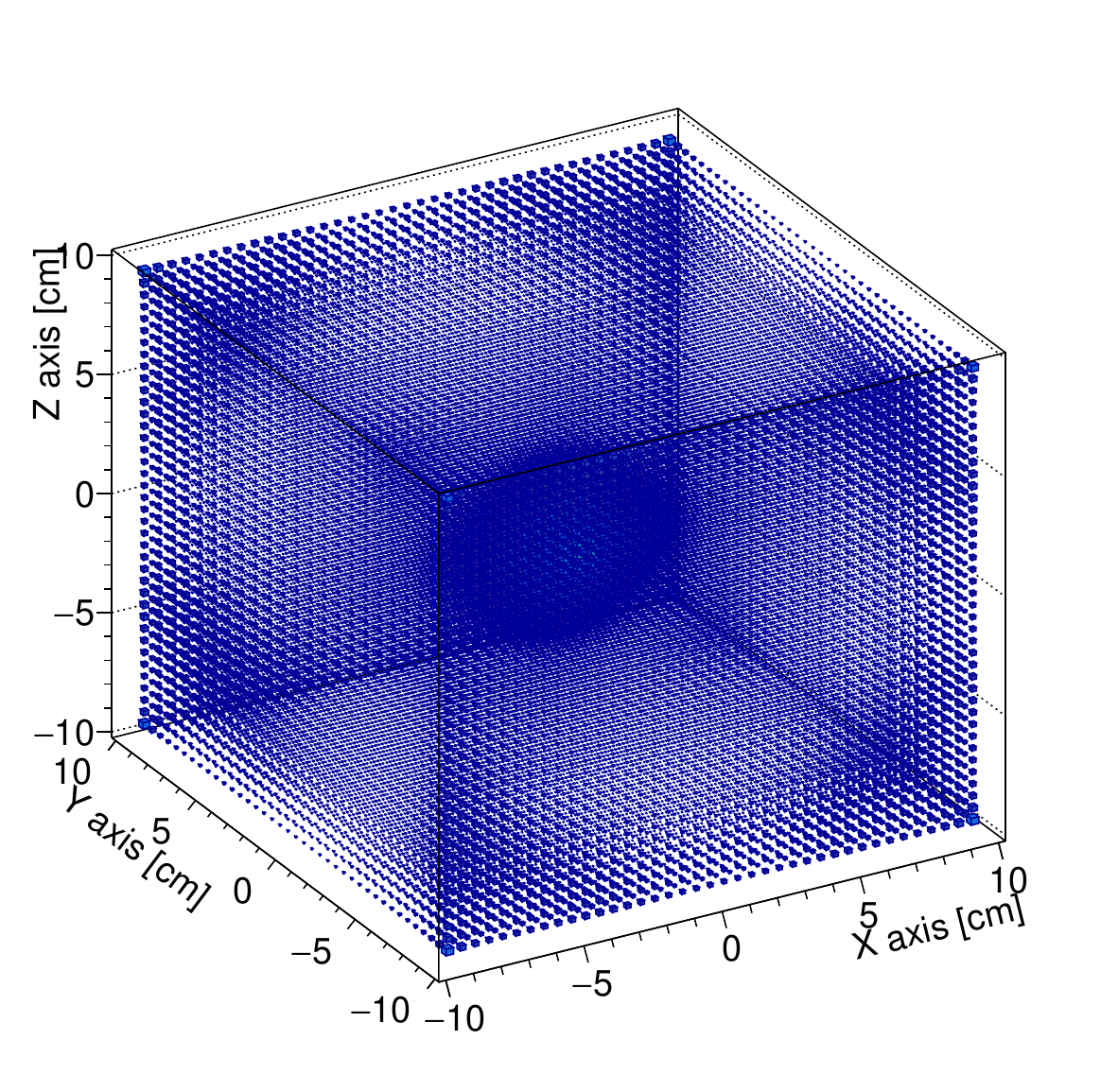}
    d)
    \includegraphics[width=0.45\textwidth]{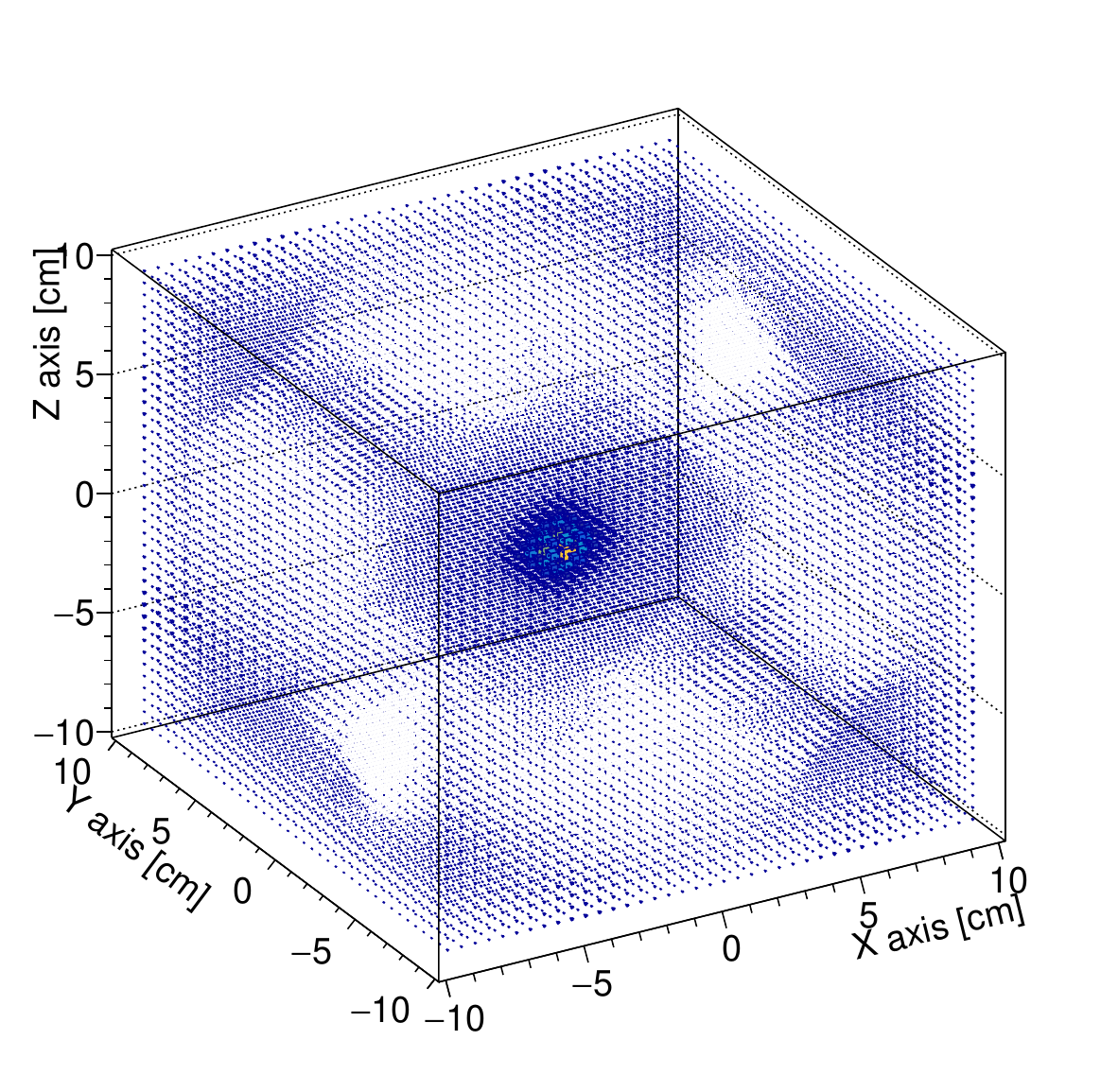}    
    \caption{Various steps of the LM-MLEM 3D reconstruction algorithm. \textit{a)} shows the reconstructed 3D image after backprojection, \textit{b) and c)} show the intermediate state after 10 and 50 iterations of LM-MLEM, respectively, and \textit{d)} shows the result after 200 iterations of LM-MLEM.}
    \label{fig:3Dsteps}
\end{figure*}

The results of applying the 3D reconstruction algorithm introduced in Sec.~\ref{sec:MandM}, based on 1) parallel-processing back-projection and 2) LM-MLEM to reproduce the 478 keV gamma-ray source distribution, is presented for several simplified cases. In these MC calculations each Compton camera geometry was optimized according to the aspects discussed in the preceding section. Figure \ref{fig:3Dsteps} shows the different steps of the 3D reconstruction for a point source located at the center. The Z axis, corresponding to the vertical axis, and the X axis, within the horizontal plane, have a pair of Compton Camera modules in front of the facets of the cube, oriented towards the origin. The remaining Y axis is left free as the principal or longitudinal axis. All figures show sizes in cm. Each Compton imager scatterer plane is located 21 cm apart from the field of view center. This distance has been chosen taking into account the size of the head from the Snyder model, which has a 11 cm major semiaxis length, and an additional distance of 10 cm from the surface (skin) to the detector.  Figure \ref{fig:3Dsteps}\textit{,a} shows the output of the back-projection step, which is subsequently used as input for the iterative LM-MLEM. Figure \ref{fig:3Dsteps}\textit{,b,c and d} show the reconstruction after 10, 50 and 200 iterations.

\begin{figure*}[h!]
    \centering
    a)
    \includegraphics[width=0.45\textwidth]{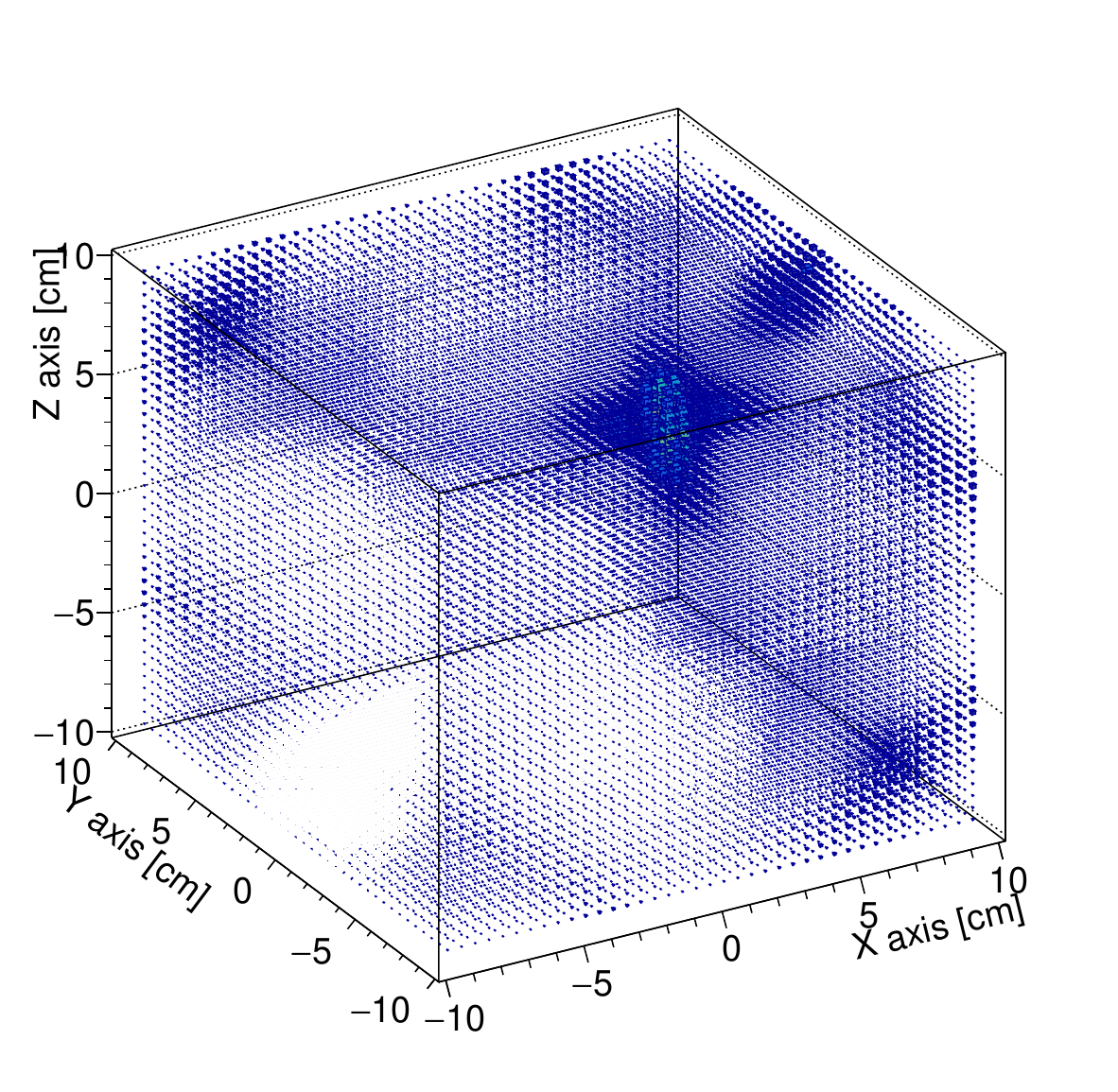}
    b)
    \includegraphics[width=0.45\textwidth]{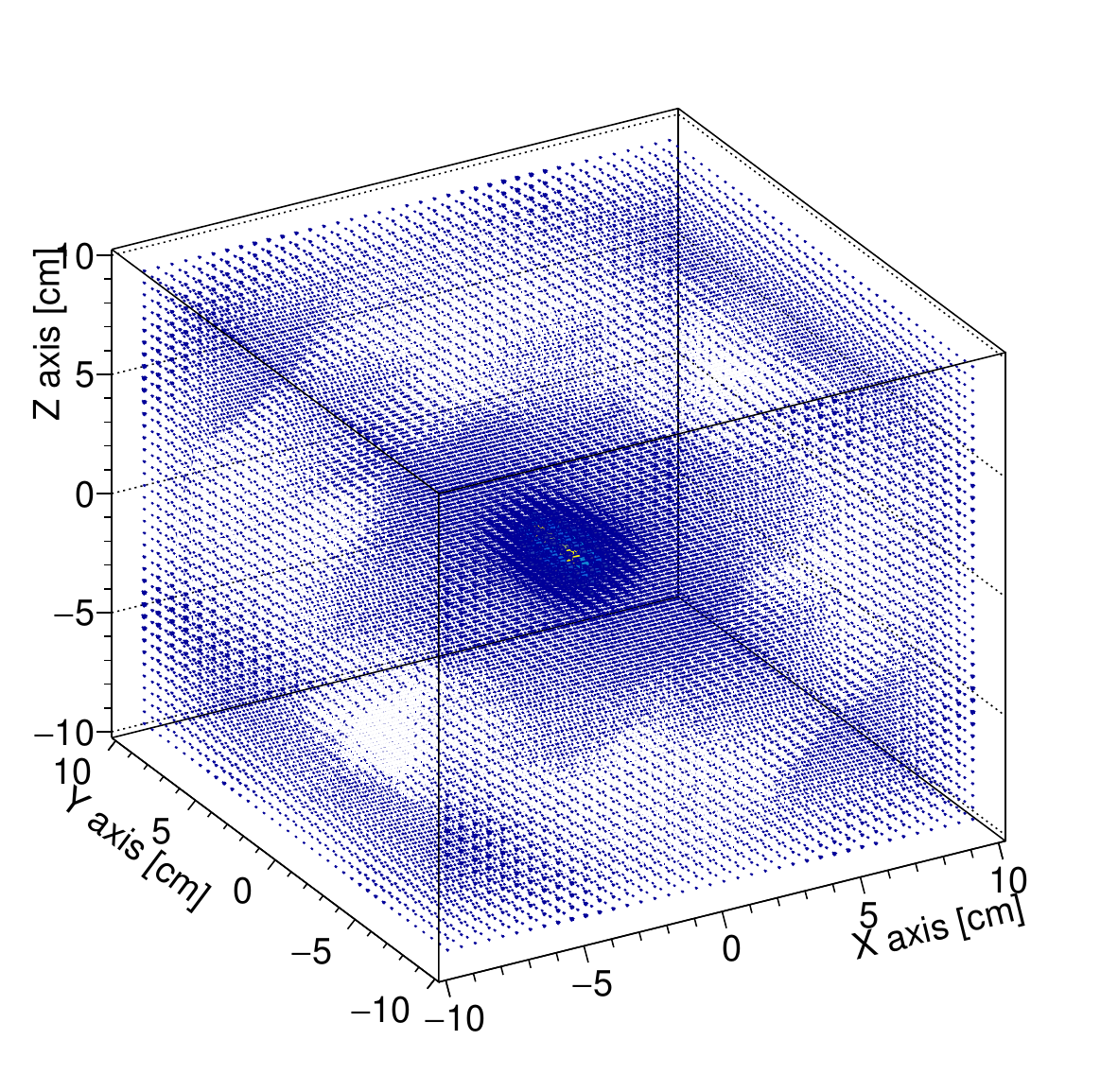}
    \\ c)
    \includegraphics[width=0.45\textwidth]{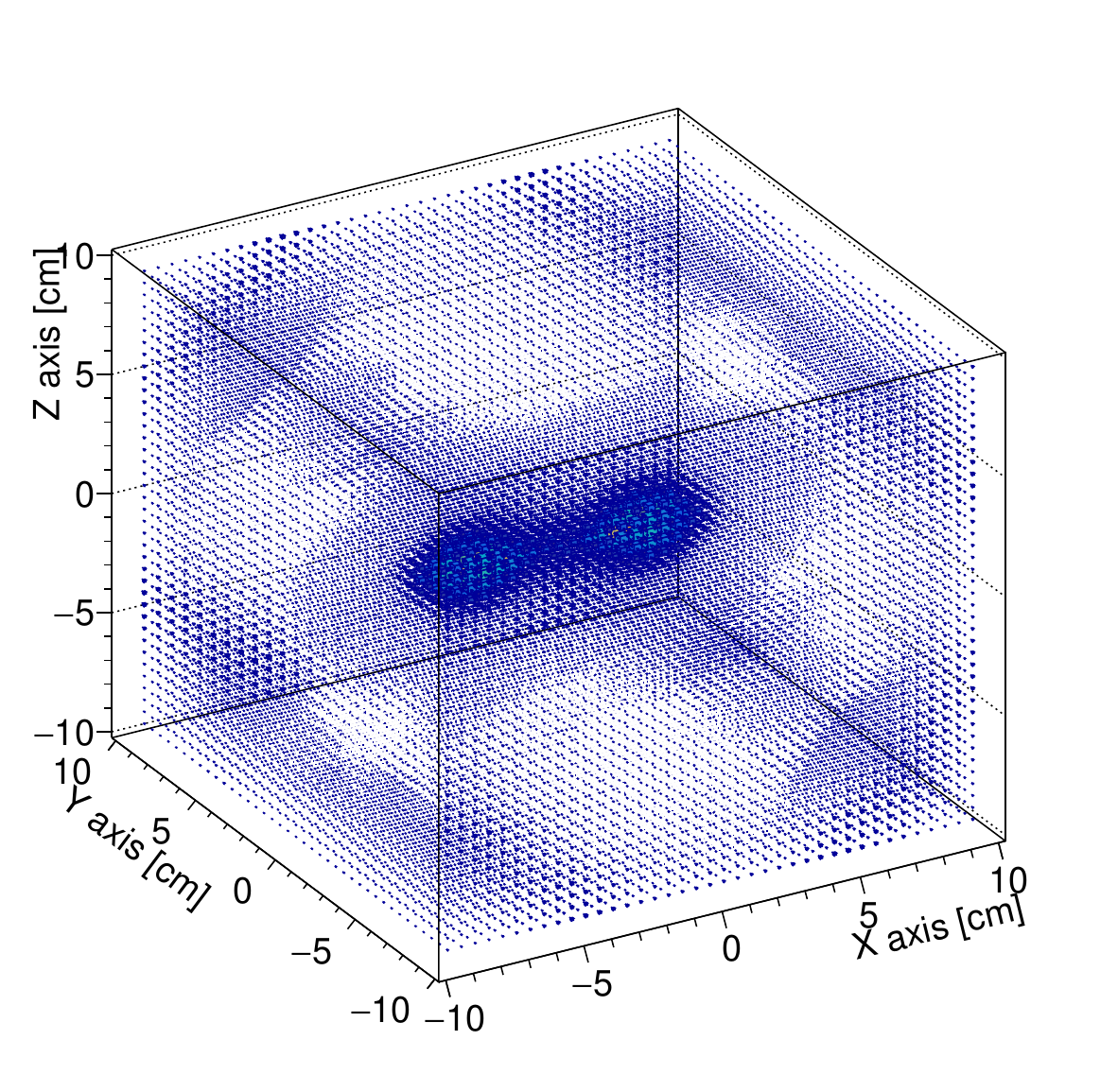}
    d)
    \includegraphics[width=0.45\textwidth]{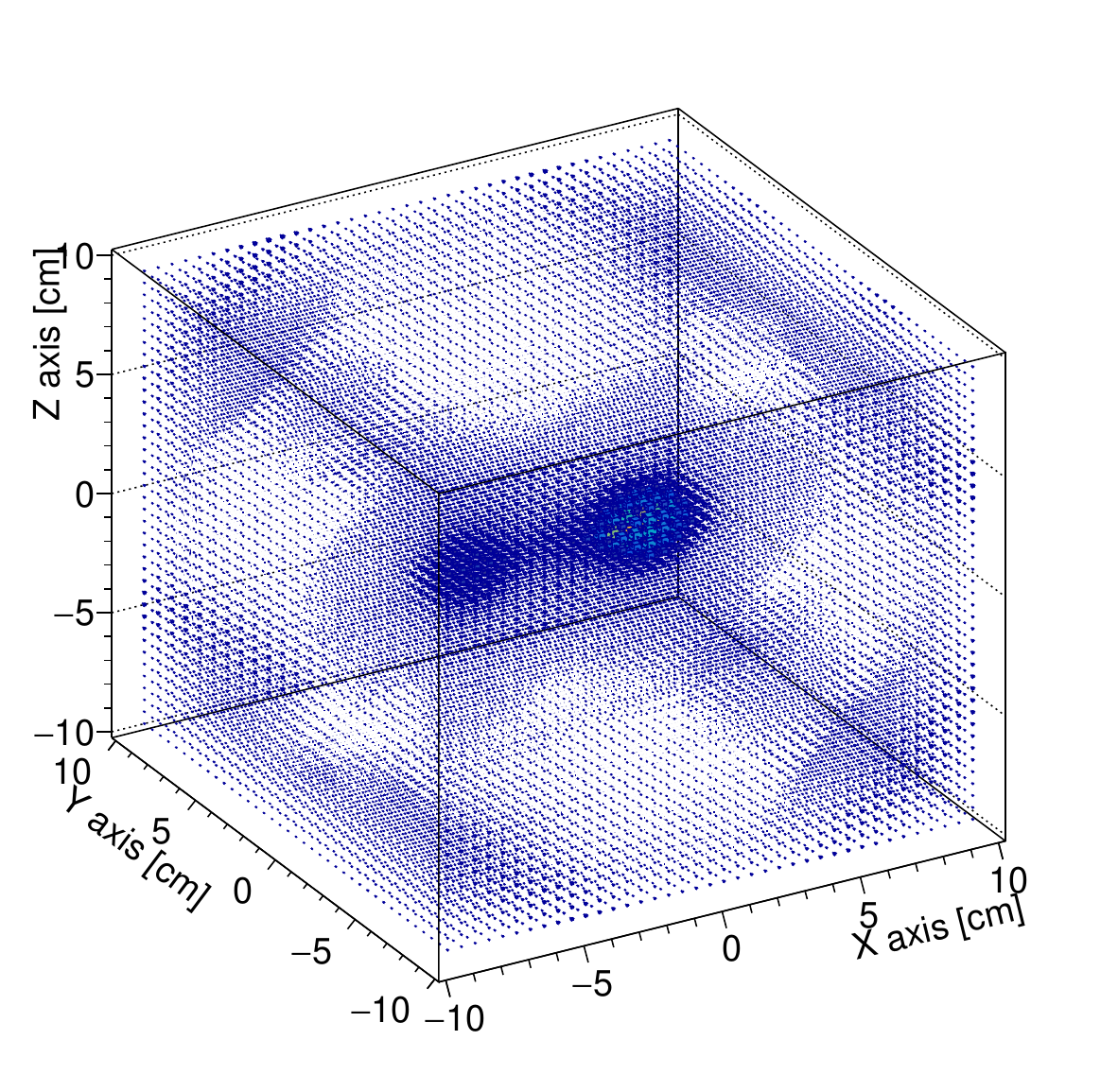}    
    \caption{ 3D reconstructed images of point-like isotropic gamma-ray sources (478 keV) after 200 iterations of the LM-MLEM algorithm. \textit{a)} shows an off-centered source, \textit{b} shows two sources in close configuration (2 cm apart near the center, located along the Y axis), \textit{c)} shows two sources located 6 cm apart along the X axis, and \textit{d} shows two sources of different intensity (2:1 ratio). }
    \label{fig:3Dexamples}
\end{figure*}

As a further validation test of the implemented 3D reconstruction methodology, a single and multiple point-like sources where placed at different positions within the Field of View of the camera array. Figure \ref{fig:3Dexamples} shows the corresponding 3D reconstructed images after 200 iterations of the LM-MLEM algorithm. Figure \ref{fig:3Dexamples}\textit{,a} shows an off-centered point source (at (3,3,3) cm). Figure \ref{fig:3Dexamples}\textit{,b} shows two-point sources of the same intensity, located along the axial axis( at (0,-1,0) and (0,1,0) cm). Figure \ref{fig:3Dexamples}\textit{,c} does the same for two-sources along the X axis (at (-3,0,0) and (3,0,0) cm). Lastly, Figure \ref{fig:3Dexamples}\textit{,d} shows two point sources of 1:2 relative intensity ratio in the same positions as c).

\section{Outlook and prospects}
\label{sec:prospects}
A first experimental campaign to test the i-TED capabilities towards BNCT was performed in October 2023 at the H22 thermal neutron beamline of  Institut Laue Langevin (ILL).  The goal of this first approximation was to check the ability of detecting the 478 keV gamma-rays from $^{10}$B captures and use it to determine the boron uptake of several different cell lines. The experiment was performed by working in parallel to the FIPPS HPGe detector array, where one of the HPGe clover detectors was replaced by one single i-TED module. Full details of the measurement and conclusions derived from this first test can be found in Ref.\cite{Lerendegui24}.

A second experimental campaign at ILL has been recently performed in June 2024, this time aiming at probing i-TED imaging capabilities in a closer-to-clinical environment, specially in terms of boron concentrations and background contamination. This time, the full i-TED array with 4 modules in cross-configuration has been used. A series of measurements with increasing background levels have been performed. The first runs consisted of dry boron-containing samples inside dishes, located at the center of the i-TED field of view. The next step included filling the dishes (8 ml in volume) with borated water at various concentration levels, resembling those of typical boron concentrations in tumor (e.g. 65 ppm). This configuration aims at having a background level typical of what is generated by a small part of a tumor.

\begin{figure}[h!]
    \centering
    \includegraphics[width=0.35\textwidth]{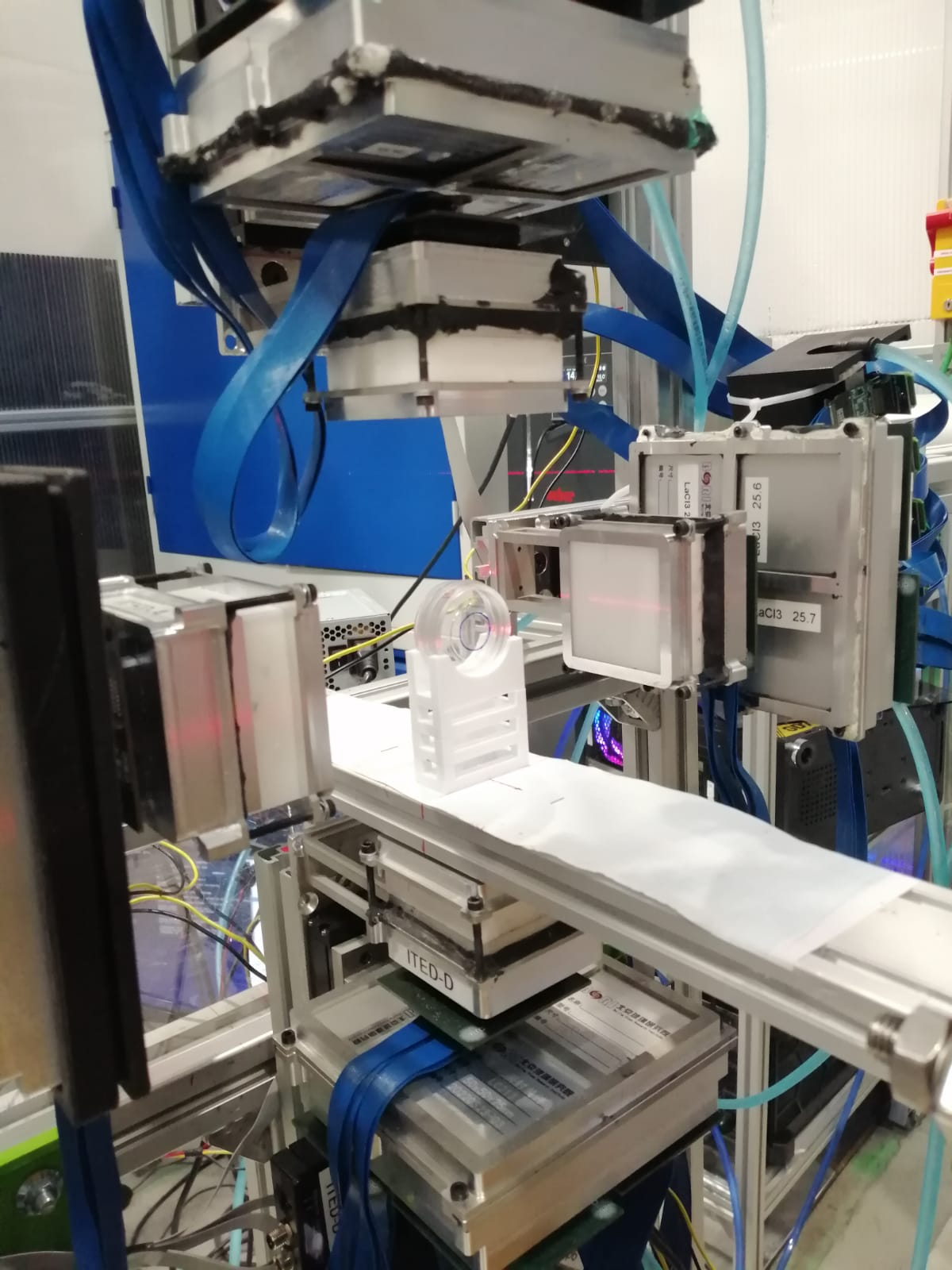}    
    \caption{Picture of the i-TED setup at ILL with a small borated-water sample (dish) in the center of the field of view of the detector array.}
    \label{fig:ILL24_Dish}
\end{figure}

The following level includes adding a phantom-like block, made of polyethylene (PE), in which a hole is drilled to place the borated-water dishes. By adding PE, the neutron captures in hydrogen strongly enhance the gamma radiation background, thus challenging the boron detection and imaging capabilities. Moreover, some imaging resolution tests were performed in this configuration, by placing two borated-water dishes inside the phantom, at different locations.

\begin{figure}[h!]
    \centering
    \includegraphics[width=0.48\textwidth]{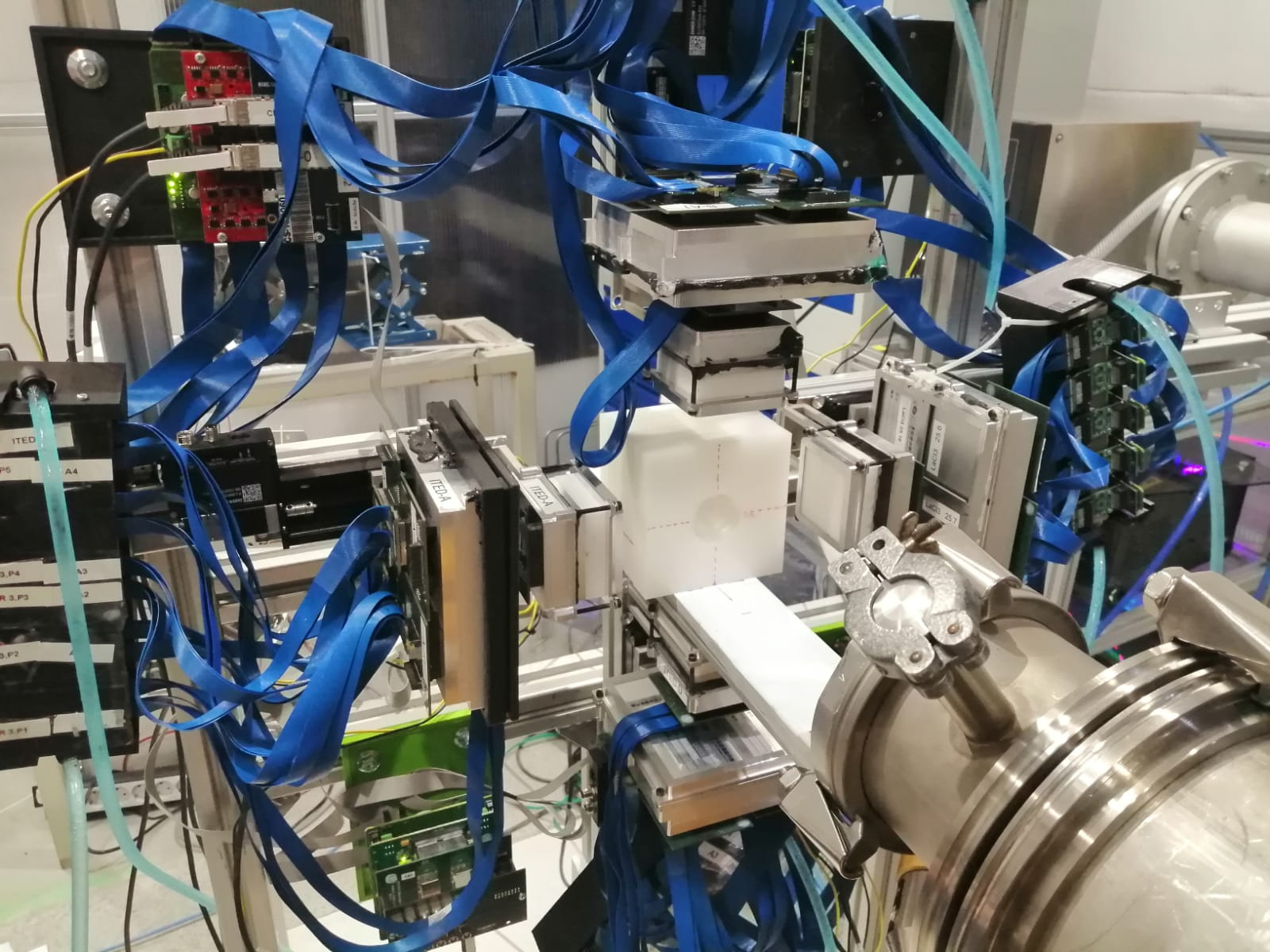}    
    \caption{Picture of the i-TED setup at ILL with a polyethylene block at the center of the field of view. The polyethylene block has a partial drill in the center to place a borated-water dish on it.}
    \label{fig:ILL24_PE}
\end{figure}

The last set of measurements at ILL targeted an even closer radiation environment for the measurements and consisted of using a water container in the i-TED field of view, first with distilled water and a borated-water dish, as a test similar to the previous ones; and lastly with also borated water at a lower concentration (20 ppm) in the container (to mimic the surrounding healthy tissues) and the borated-water dish (100 ppm) as tumor-like volume. This last test also served as a resolution plus contrast test, to validate the i-TED imaging capabilities towards BNCT.

\begin{figure}[h!]
    \centering
    \includegraphics[width=0.48\textwidth]{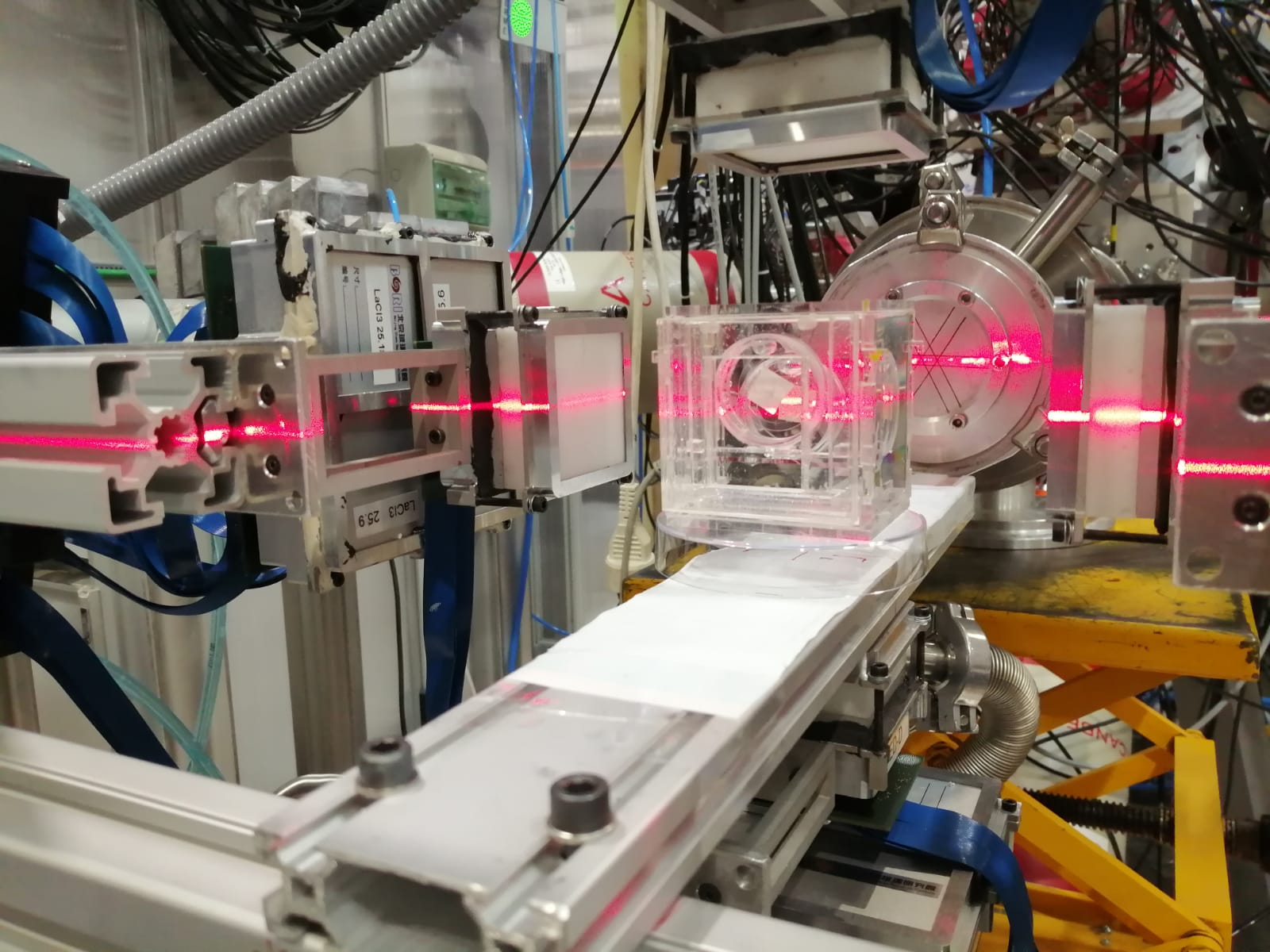}    
    \caption{Picture of the i-TED setup at ILL with a water container at the center of the field of view. One or several borated-water dishes (two in the picture) are placed inside the container. }
    \label{fig:ILL24_Water}
\end{figure}

\section{Conclusions}
\label{sec:Conclusions}

This study has explored the potential of the i-TED Compton imaging system, initially developed for nuclear astrophysics, for real-time dosimetry in Boron Neutron Capture Therapy (BNCT).  The results indicate that optimizing the detector configuration significantly enhances its ability to detect 478 keV gamma rays, crucial for accurate boron distribution imaging, with special consideration to the harsh radiation environment of a BNCT treatment. The developed 3D reconstruction algorithm has been tested with simulated point-like sources with expected count rates similar of a treatment, showing promise for real-time treatment monitoring. Preliminary experiments have been performed to validate the capabilities of the i-TED array, laying the groundwork for its further development and clinical application in BNCT. Future work will focus on refining these technologies and conducting comprehensive clinical validations to ensure their efficacy in improving BNCT dosimetry.

\section*{Declaration of competing interest}
The authors declare that they have no known competing financial interests or personal relationships that could have appeared to influence the work reported in this paper.

\section*{CRediT authorship contribution statement}
\textbf{P. Torres-S\'anchez}: Investigation, Methodology, Formal analysis, Visualization, software, Writing - original draft.
\textbf{J. Lerendegui-Marco:} Investigation, Methodology, Formal analysis, Writing-review \& editing.
\textbf{J. Balibrea-Correa:} Investigation, Methodology, software.
\textbf{V. Babiano-Su\'arez:} Investigation, Methodology, software.
\textbf{B.~Gameiro:} Investigation, Methodology, software. 
\textbf{I. Ladarescu:} Investigation, Software. 
\textbf{P.~\'Alvarez-Rodr\'iguez:} Investigation.
\textbf{J.M. Daugas}: Investigation, Methodology.
\textbf{U. Koester}: Investigation, Methodology.
\textbf{C. Michelagnoli}: Investigation, Resources, Writing - review \& editing. 
\textbf{M. Pedrosa-Rivera}: Investigation.
\textbf{I. Porras}: Conceptualization, Investigation, Resources, Project administration, Funding acquisition, Writing -review \& editing.
\textbf{M.C. Ruiz-Magaña}: Investigation.
\textbf{C. Ruiz-Ruiz}: Investigation, Resources.
\textbf{C. Domingo-Pardo:} Conceptualization, Investigation, Methodology, Supervision, Project administration, Funding acquisition, Writing - review \& editing.

\section{Acknowledgements}

This work builds upon research conducted under the ERC Consolidator Grant project HYMNS (grant agreement No. 681740) and has been supported by the ERC Proof-of-Concept Grant project AMA (grant agreement No. 101137646). We also acknowledge funding from the Spanish Ministerio de Ciencia e Innovación under grants PID2022-138297NB-C21 and PID2019-104714GB-C21, as well as from CSIC under grant CSIC-2023-AEP128, and Fundación Científica AECC under grant INNOV223579PORR. Additionally, the authors thank the support provided by postdoctoral grants FJC2020-044688-I and ICJ220-045122-I, funded by MCIN/AEI/10.13039/501100011033 and the European Union NextGenerationEU/PRTR;  postdoctoral grant CIAPOS/2022/020 funded by the Generalitat Valenciana and the European Social Fund and a PhD grant PRE2023 from CSIC. Financial support from the Institut Laue-Langevin during the experimental campaign is also gratefully acknowledged.

 \bibliographystyle{elsarticle-num} 





\end{document}